\documentclass[conference]{IEEEtran}
\IEEEoverridecommandlockouts

\usepackage{cite}
\usepackage{amsmath,amssymb,amsfonts}
\usepackage{graphicx}
\usepackage{textcomp}
\usepackage{xcolor}
\usepackage{array}
\usepackage[caption=false,font=normalsize,labelfont=sf,textfont=sf]{subfig}
\usepackage{url}
\usepackage[colorlinks=true, linkcolor=blue, citecolor=blue, urlcolor=blue, bookmarks=false, pdfpagemode=UseNone]{hyperref}
\usepackage{makecell}
\usepackage{enumitem}
\usepackage[linesnumbered,ruled]{algorithm2e}
\usepackage{booktabs}
\def\BibTeX{{\rm B\kern-.05em{\sc i\kern-.025em b}\kern-.08em
    T\kern-.1667em\lower.7ex\hbox{E}\kern-.125emX}}
\usepackage{fancyhdr}

\fancypagestyle{globecom}{%
	\fancyhf{}
	\fancyhead[L]{\footnotesize\leftmark}
	\fancyhead[R]{\footnotesize\thepage}
	
}

\makeatletter

\def\ps@IEEEtitlepagestyle{%
	\def\@oddhead{%
		\hfil
		\footnotesize
		\leftmark
		\hfil
	}%
	\def\@evenhead{%
		\hfil
		\footnotesize
		\leftmark
		\hfil
	}%
	\def\@oddfoot{%
		\parbox{\textwidth}{%
			\centering
			\scriptsize
			\copyright{} 2026 IEEE.
			Personal use of this material is permitted.
			Permission from IEEE must be obtained for all other uses,
			in any current or future media, including reprinting/republishing
			this material for advertising or promotional purposes, creating new
			collective works, for resale or redistribution to servers or lists,
			or reuse of any copyrighted component of this work in other works.
		}%
	}%
	\def\@evenfoot{%
		\@oddfoot
	}%
}

\begin{document}
\setlength{\columnsep}{0.25in}

\title{Cross-Domain Joint DDoS Detection in Multi-Controller SDN via Confidence-Based Entropy Fusion
\thanks{This work is supported by the National Natural Science Foundations of China under Grants 62203062 and 61932005, and partly by the Fundamental Research Funds for the Central Universities under Grant 2242022k60006. Corresponding author: Shen Wang (e-mail: shen.wang@bupt.edu.cn).}
}

\author{%
\IEEEauthorblockN{Zhaoyang Zhang, Shen Wang, and Xiaofeng Tao}
\IEEEauthorblockA{%
National Engineering Research Center for Mobile Network Technologies,\\
Beijing University of Posts and Telecommunications, Beijing, China, 100876\\
Email: \{zhangzhaoyang, shen.wang, taoxf\}@bupt.edu.cn}
}

\maketitle

\markboth
	  {To Appear in IEEE Global Communications Conference, Macau S.A.R., China, 7 -- 11 December, 2026}
	  {}

\begin{abstract}

In multi-controller Software-Defined Networking (SDN), Distributed Denial-of-Service (DDoS) attacks exhibit a ``dispersed source, concentrated target'' pattern across domains, i.e., attack traffic originates from multiple edge-controller domains but converges on a victim in a single aggregation controller domain. While entropy-based DDoS detectors are effective in single-controller settings, their direct application in multi-controller SDN reveals a previously overlooked anomaly. Through systematic experiments, we identify an \emph{aggregation bias}: during the post-attack transition phase, the aggregation controller continues to generate excessive false positives, while edge controllers have already returned to normal. We attribute this phenomenon to the coupled effects of OpenFlow statistics lag and unconstrained dynamic-threshold drift. To address this issue, we propose a cross-domain confidence-fusion framework that leverages lightweight edge-side messages to calibrate aggregation-controller decisions without sharing raw traffic data. The framework is non-intrusive, communication-efficient, and incrementally deployable. Experiments on a three-controller linear Mininet testbed with 24 hosts over 10 runs show that the method preserves edge-controller performance while reducing the aggregation false positive rate from 8.87\% to 1.96\% and increasing the F1 score from 89.04\% to 96.89\%.

\end{abstract}


\section{Introduction}\label{sec:intro}


Software-Defined Networking (SDN) has become a cornerstone of modern network infrastructure, owing to its decoupling of control and data planes and its global programmability~\cite{openflow}. To overcome the scalability bottlenecks of single-controller architectures, distributed multi-controller deployments have become the mainstream solution~\cite{multi_controller_survey}, partitioning the network into multiple control domains.

\begin{figure}[!htbp]
	\centering
	\vspace{-0.7em}
	\includegraphics[width=0.92\columnwidth]{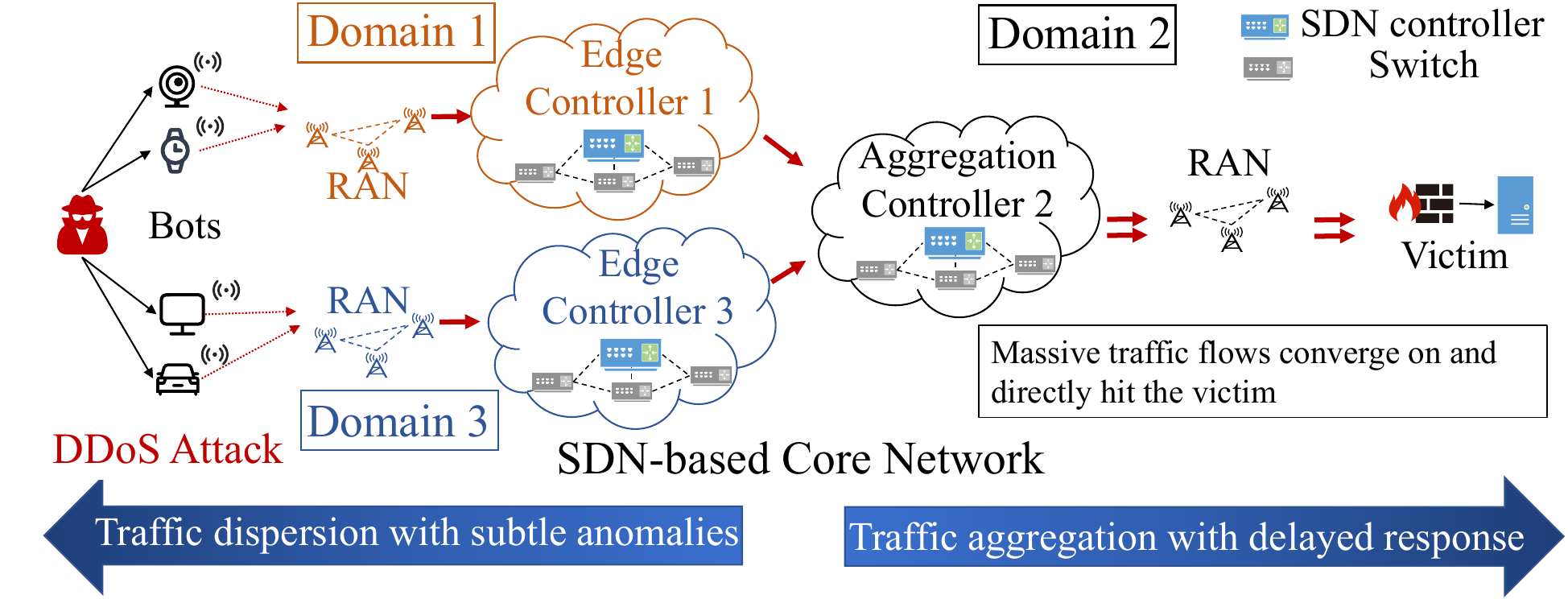}
	\caption{Cross-domain DDoS attack scenario. Attack traffic is dispersed in edge domains and converges at the victim within the aggregation domain.}
	\label{fig:ddos_scenario}
	\vspace{-0.7em}
\end{figure}

However, this decentralization also poses new challenges to network security. Distributed Denial-of-Service (DDoS) attacks~\cite{chahal2024ddos} are typically characterized by a ``highly dispersed source, concentrated target'' pattern: attack traffic, generated by multiple compromised hosts (bots), is dispersed across multiple control domains and converges in the aggregation domain hosting the victim (see Fig.~\ref{fig:ddos_scenario}). For exposition, we call controllers whose domains attach these attack-source hosts \emph{edge controllers}, and those attaching the victim \emph{aggregation controllers}; prior works use near-synonyms such as ``source/destination''~\cite{yu2021cooperative} or ``local/central''~\cite{tayfour2021collaborative}.

Entropy-based DDoS detection achieves favorable performance in single-controller settings, owing to its low overhead and strong interpretability~\cite{JESS2018,Entropy_2022_TNSM}. A natural engineering intuition is that deploying such a detector independently on every controller such as edge and aggregation controllers in Fig.~\ref{fig:ddos_scenario} should leverage richer distributed observations and thus further improve effectiveness.

However, our systematic experiments reveal a \emph{counter-intuitive phenomenon}: when such detectors are directly reused, the aggregation controller's performance \emph{degrades rather than improves}, with three salient features: (i)~spatially, the detection bias concentrates on the aggregation controller, while edge controllers remain essentially unaffected; (ii)~temporally, it concentrates in the post-attack transition phase, with in-attack detection being almost perfect; and (iii)~within the same windows, edge controllers correctly return to normal state while the aggregation controller still alarms. We term this {systematic detection bias} the \emph{aggregation bias}. The central question is: \emph{without sharing raw traffic data, how can we exploit limited cross-domain information to eliminate this bias and achieve consistent, robust global DDoS detection?}

Existing SDN DDoS detection studies fall into three categories. (i)~{Statistics- and entropy-based detection}~\cite{JESS2018,Entropy_2022_TNSM,Generalized_Entropy_SAHOO_2018} characterizes anomalies via Shannon entropy, generalized entropy, or information distance; lightweight and interpretable, but assumes a single observation point with a globally consistent view. (ii)~{Machine-learning-based detection} trains classifiers over flow-level features but requires large labeled datasets and adopts a single-controller perspective without modeling inter-controller fusion. (iii)~{Multi-controller cooperative detection}~\cite{giotis2014leveraging,Entropy_dstIP_2015,yu2021cooperative,tayfour2021collaborative,wang2023ccguard} explores cooperative flow-statistics sampling, distributed entropy detection, two-stage edge--center coordination, rule sharing, and cross-domain protection. These works focus on \emph{how to share information among multiple controllers} and assume comparable observation quality at every controller; they neither characterize the spatial asymmetry of the ``dispersed source, concentrated target'' pattern, nor identify the systematic aggregation-side bias caused by statistics lag and threshold drift.

Building on a mechanism-level analysis of the aggregation bias, this paper proposes a lightweight, incrementally deployable cross-domain confidence-fusion framework that, without sharing raw traffic, leverages edge-side information to constrain and calibrate the aggregation controller's decisions. The main contributions of this paper are as follows:

\begin{itemize}
	\item  This paper identifies a previously overlooked phenomenon, termed \emph{aggregation bias}, when entropy-based DDoS detection method is deployed directly on each individual controller in an SDN multi-controller setup, and explains it through two coupled root causes, which are interpreted as the interaction between local observation distortion and the lack of cross-domain calibration.
	


	\item A cross-domain confidence-fusion joint detection method is proposed. Built on a confidence metric and per-source edge consensus, the aggregation controller's decision is upgraded from a purely local binary judgment to a reliable inference calibrated by cross-domain information.
	
	\item This paper develops a non-intrusive deployment mechanism with robust fallback capability. The method relies only on lightweight scalar messages from edge controllers and degrades to local detection when reports are absent. Experiments show that the aggregation false positive rate drops from 8.87\% to 1.96\% while edge-controller performance is unchanged.
 
\end{itemize}

The rest of this manuscript is organized as follows. System overview and problem formulation is presented in Section~\ref{sec:model}. Section~\ref{sec:bias} presents aggregation-bias phenomenon and analyzes bias mechanisms. A cross-domain confidence-fusion joint detection method is proposed in Section~\ref{sec:method}. Section~\ref{sec:experiment} illustrates main results via experiments. Section~\ref{sec:conclusion} concludes.

\section{System Overview and Problem Formulation}\label{sec:model}

\subsection{System Architecture and Assumptions}\label{subsec:architecture}

\begin{figure}[!htbp]
    \centering
    	\vspace{-0.8em}
    \includegraphics[width=0.85\columnwidth]{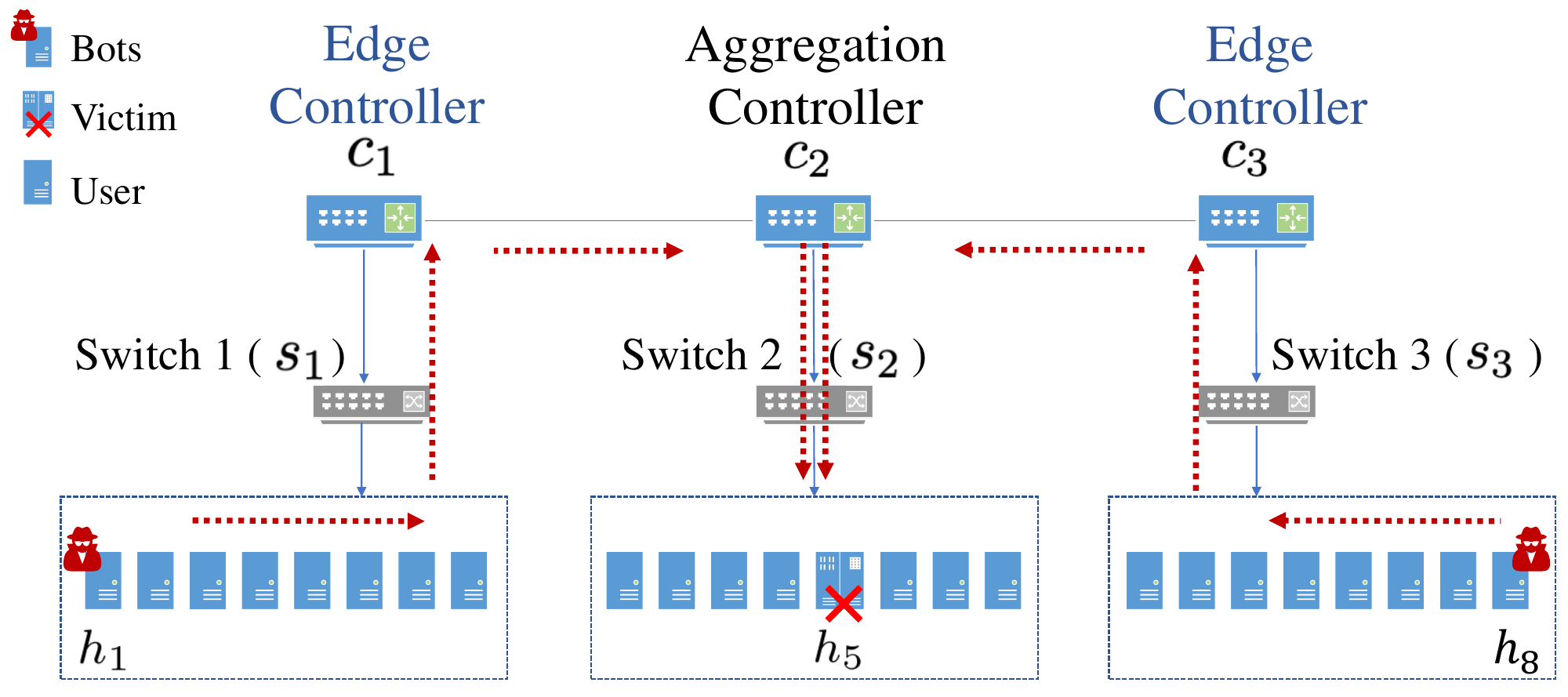}
    \caption{Three-controller SDN testbed topology. Controllers $c_1$ and $c_3$ act as edge controllers, while $c_2$ serves as the aggregation controller. Compromised hosts (bots) $h_1$ and $h_8$ generate DDoS traffic targeting the victim $h_5$.}
    \label{fig:topo_multi}
    	\vspace{-0.8em}
\end{figure}

To validate the proposed cross-domain joint DDoS detection framework, we adopt the three-controller topology shown in Fig.~\ref{fig:topo_multi}, which abstracts the cross-domain attack scenario in Fig.~\ref{fig:ddos_scenario}. Edge controllers $c_1$ and $c_3$, together with the aggregation controller $c_2$, manage a linear switch chain $s_1$--$s_2$--$s_3$, with 24 hosts evenly distributed across the three domains (i.e., 8 per domain). 

Compromised hosts (bots) $h_1$ in the $c_1$ domain and $h_8$ in the $c_3$ domain act as attack sources, targeting the victim $h_5$ in the $c_2$ domain. All cross-domain traffic is forwarded via switch $s_2$. Each controller periodically collects intra-domain flow statistics through OpenFlow and executes a local entropy-based detector. In addition, the aggregation controller $c_2$ performs cross-domain confidence fusion and outputs the global decision $\hat{y}$. Edge controllers transmit only distilled scalar metrics (without raw traffic) to $c_2$ through edge-to-aggregation application-layer reports over TCP, while $c_2$ sends no application-layer report back.

We assume: \textbf{(A1)} only distilled scalar metrics are shared edge-to-aggregation, which limits communication overhead and avoids privacy issues of raw-traffic exchange; \textbf{(A2)} inter-controller links are reliable with latency much smaller than the 1-s OpenFlow polling period---this holds in local data-center environments; \textbf{(A3)} the attack path traverses at least one edge and one aggregation domain (canonical cross-domain DDoS); \textbf{(A4)} controllers themselves are trusted.

Compromised controllers, tampered, replayed, or reordered reports, and inter-controller delays violating assumption~\textbf{(A2)} are outside the current model.

\subsection{Formal Model}\label{subsec:formal_model}

Consider a distributed SDN governed by $M$ controllers $\mathcal{C}=\{c_1,\ldots,c_M\}$, where each $c_i$ governs a domain $\mathcal{S}_i$ of OpenFlow switches with $\mathcal{S}_i\cap\mathcal{S}_j=\emptyset$ for $i\neq j$. Under the ``dispersed source, concentrated target'' DDoS attack pattern, $\mathcal{C}$ is partitioned into edge controllers $\mathcal{C}_{\mathrm{edge}}\subset\mathcal{C}$ (domains attaching attack-source hosts) and aggregation controllers $\mathcal{C}_{\mathrm{agg}}\subset\mathcal{C}$ (domains attaching the victim), with $\mathcal{C}_{\mathrm{edge}}\cup\mathcal{C}_{\mathrm{agg}}=\mathcal{C}$ and $\mathcal{C}_{\mathrm{edge}}\cap\mathcal{C}_{\mathrm{agg}}=\emptyset$. In our testbed (see Fig.~\ref{fig:topo_multi}), $\mathcal{C}_{\mathrm{edge}}=\{c_1,c_3\}$ and $\mathcal{C}_{\mathrm{agg}}=\{c_2\}$.

Each controller $c_i \in \mathcal{C}_{\mathrm{edge}}\cup\mathcal{C}_{\mathrm{agg}}$ uses a sliding window of $N$ packets, indexed by $n=1,2,\ldots$ in chronological order, and outputs a normalized \emph{entropy metric} $H_{i}(n)\in[0,1]$ derived from the Shannon entropy~\cite{shannon1948mathematical} of discriminative traffic attributes (e.g., destination IP/port)~\cite{JESS2018,Entropy_2022_TNSM}; that is, the full construction follows a single-controller entropy-based detector~\cite{JESS2018,zhang2026timemeetsspaceentropy} to detect DDoS attack. Ideally, entropy metric $H_{i}(n)$ should faithfully reflect the ground truth state $s(n)\in\{0,1\}$ (0: normal, 1: attack) from traffic generator, but in multi-controller settings polling latency and feature-selection mismatches induce a role-dependent distortion $H_{i}(n) = \tilde{H}_{i}(n) + \varepsilon_i(n)$, where $\tilde{H}_{i}(n)$ is an unbiased ideal entropy observation, and $\varepsilon_i(n)$ is modeled as a zero-mean but role-dependent stochastic distortion/bias between the ideal entropy observation $\tilde H_{i}(n)$ and the measured value $H_{i}(n)$. Each SDN controller $c_i \in \mathcal{C}_{\mathrm{edge}}\cup\mathcal{C}_{\mathrm{agg}}$ in a domain then compares $H_{i}(n)$ against a dynamic threshold $H_{i}^{\mathrm{thd}}(n)$, Exponentially Weighted Moving Average (EWMA-) tracked over slow variations of background traffic~\cite{JESS2018,zhang2026timemeetsspaceentropy}, and outputs a binary local detection decision  
 \begin{equation}
	y_i(n)=
	\begin{cases}
		1, & H_{i}(n)<H_{i}^{\mathrm{thd}}(n), \\[4pt]
		0, & H_{i}(n)\ge H_{i}^{\mathrm{thd}}(n).
	\end{cases}
	\label{eq:local_rule}
\end{equation}

\subsubsection{Aggregation bias}

In the post-attack transition phase (i.e., $s(n)=0$), if
\begin{equation}
	\begin{aligned}
		&\mathbb{P}\!\big(y_j(n)=1 \mid s(n)=0\big) \gg \mathbb{P}\!\big(y_i(n)=1 \mid s(n)=0\big), \\
		&\quad \forall c_j \in\mathcal{C}_{\mathrm{agg}},\, c_i \in\mathcal{C}_{\mathrm{edge}},
	\end{aligned}
	\label{eq:bias_def}
\end{equation}
we say the system exhibits an \emph{aggregation bias}, meaning that aggregation controllers are significantly more likely than edge controllers to raise false attack alarms after the attack has already ceased. Empirically, aggregation domains exhibit larger distortion magnitudes, i.e.,$| \varepsilon_j(n) | > | \varepsilon_i(n) |, j\in\mathcal{C}_{\mathrm{agg}},\, i\in\mathcal{C}_{\mathrm{edge}}$ mainly due to delayed flow-statistics updates. This distorted observation is further amplified by the EWMA threshold self-update, causing persistent false positives at aggregation controllers. Section~\ref{sec:bias} quantifies this effect and traces it to two coupled root causes.

\subsubsection{Cross-domain fusion problem} 

To mitigate the bias, the aggregation controller $c_j$ receives lightweight reports $m_i(n)$ from each $i\in\mathcal{C}_{\mathrm{edge}}$ and produces a fused DDoS detection decision
\begin{equation}
\begin{aligned}
    \hat{y}(n) = \mathcal{F}\!\big(&H_{j}(n),\, H_{j}(n)^{\mathrm{thd}},\{m_i(n)\}_{c_i\in\mathcal{C}_{\mathrm{edge}}}\big),
\end{aligned}
\label{eq:fusion_obj}
\end{equation}
subject to: (i)~$m_i(n)$ is low-dimensional scalar metrics/information; (ii)~edge-controller logic is unchanged (zero intrusion); and (iii)~cross-domain fusion design $\mathcal{F}$ degrades to the local decision when reports are absent. Section~\ref{sec:method} proposes the key $m_i(n)$ and $\mathcal{F}$.

\section{Aggregation Bias: Phenomenon and Mechanism}\label{sec:bias}

\subsection{Baseline Observation}\label{subsec:baseline}
To illustrate the aggregation bias phenomenon, we evaluate the local DDoS detector (Equation~\eqref{eq:local_rule}) independently on all three controllers by replaying the raw packet windows from 10 testbed trials (Fig.~\ref{fig:topo_multi}) without cross-domain information or active cross-domain rules. Although the aggregation controller $c_2$ operates at the cross-domain forwarding hub, its observations are still limited to local statistics within $\mathcal{S}_2$. Table~\ref{tab:baseline} summarizes the averaged results.

Three key phenomena are observed. (i)~\emph{Spatially}, the bias is concentrated at the aggregation controller $c_2$, whose false positive rate (FPR) reaches $8.87\%$, i.e., approximately $3$--$10$ times higher than those of the edge controllers $c_1$ and $c_3$. (ii)~\emph{Temporally}, $c_2$ achieves a recall of $99.79\%$ during the attack, while nearly all of its 57.4 false-positive (FP) windows are clustered within a few consecutive windows after the attack terminates. (iii)~\emph{Cross-controller inconsistency}, during these same intervals, edge controllers $c_1$ and $c_3$ have already correctly reverted to the normal state, while the aggregation controller $c_2$ continues to raise alarms. These results indicate that the bias is not due to random fluctuations, but rather a systematic effect induced by the aggregation controller's distinct observation conditions.

 \begin{table}[!htbp]
	\centering
	\caption{Baseline DDoS detection results (local detector replay on the same 10 trials).}
	\label{tab:baseline}
	\renewcommand\arraystretch{1.1}
	\begin{tabular}{lcccc}
		\toprule
		\textit{Controllers} & \textit{Accuracy} & \textit{F1} & \textit{FPR} & Avg.\ FP \\
		\midrule
		$c_1$ (edge) & 98.84\% & 98.84\% & 0.93\% & 1.7 \\
		$c_2$ (aggregation) & 93.46\% & 89.04\% & 8.87\% & 57.4 \\
		$c_3$ (edge) & 98.40\% & 98.53\% & 3.33\% & 5.8 \\
		\bottomrule
	\end{tabular}
	\vspace{-2em}
\end{table}

\subsection{Two Coupled Root Causes}\label{subsec:causes}
\subsubsection{Root Cause I: entropy distortion under statistics lag}
The local entropy detector relies on a rate-triggered dynamic feature-selection mechanism~\cite{zhang2026timemeetsspaceentropy}: when the observed packet rate exceeds a preset threshold $\theta^{\mathrm{thd}}$, an anomaly-oriented feature-selection procedure is activated. Because $s_2$ carries all cross-domain traffic and OpenFlow statistics are polled at ${\sim}1$\,s, after the attack stops the queued packets cause the observed rate to decay in stages (a typical sequence: $16{,}520 \to 10{,}978 \to 5{,}703 \to 3{,}520$\,pps), keeping the rate above $\theta^{\mathrm{thd}}$ for several consecutive windows. The feature-selection procedure is then mistakenly applied to traffic that has already returned to normal, depressing the entropy from a normal level of ${\sim}0.7$ down to $0.10$--$0.29$ and producing immediate false alarms. In single-controller topologies the same effect spans only $1$--$2$ windows; the aggregation switch's heavier queue accumulation amplifies both the count and the intensity of affected windows.

\subsubsection{Root Cause II: unconstrained EWMA threshold drift} EWMA-based dynamic thresholding, a widely adopted adaptive mechanism in entropy-based DDoS detection~\cite{zhang2026timemeetsspaceentropy}, follows slow background variations through a self-loop on the controller's own historical entropy. In multi-controller settings this design exhibits a generic degenerate behavior. When an early attack window falls into EWMA's low-rate branch, its low entropy is absorbed into the update and the threshold enters a sustained decay sequence (e.g., $0.6 \to 0.54 \to 0.49 \to \cdots$); subsequent normal windows are misclassified, while the traffic-adaptive freezing constraint blocks recovery, forming a vicious cycle in which the threshold can collapse to near zero. Root Cause II couples temporally with Root Cause I: Root Cause I depresses entropy after the attack, then Root Cause II locks the threshold low, jointly sustaining persistent false alarms during the transition phase.

In essence, both causes reflect the same tension---the aggregation controller's local view is distorted during the transition phase, yet its EWMA threshold has no external calibrator. Edge controllers $c_1, c_3$, free of queue accumulation, recover promptly and provide exactly the two complementary pieces of information $c_2$ lacks: whether the attack has ended, and at what level the normal-traffic entropy should sit. This information complementarity motivates the cross-domain fusion design in Section~\ref{sec:method}.

\section{Cross-Domain Confidence-Fusion Method}\label{sec:method}
We instantiate the cross-domain fusion design Equation $\mathcal{F}$ in~\eqref{eq:fusion_obj} via four coupled components, all running on the aggregation controller and leaving edge-controller logic untouched: (i)~edge-to-aggregation application-layer reporting and a local confidence metric; (ii)~per-source edge-report tracking and consensus aggregation; (iii)~consensus-referenced cross-domain fusion; and (iv)~the joint-detection algorithm.

\subsection{Edge-to-Aggregation Reporting and Confidence Metric}\label{subsec:confidence}

In the proposed method, every $c_i\in\mathcal{C}_{\mathrm{edge}}$ sends the low-dimensional message $m_i(n)$ in Equation~\eqref{eq:fusion_obj} to $c_j\in\mathcal{C}_{\mathrm{agg}}$ over a TCP socket at the end of each detection window; $c_j$ sends no application-layer report back. The message carries two scalars: the local threshold $H_{i}^{\mathrm{thd}}(n)$ and a continuous detection confidence $\alpha_i(n)$ introduced next. The measured communication and processing overhead is reported in Section~\ref{subsec:overhead}.

Reporting only the binary decision $y_i$~\cite{giotis2014leveraging,Entropy_dstIP_2015,yu2021cooperative,tayfour2021collaborative,wang2023ccguard} cannot distinguish high-confidence attacks from boundary noise. We therefore lift $y_i$ into a continuous \emph{detection confidence} $\alpha_i\in[0,1]$. When $y_i(n)=0$, we set $\alpha_i(n)=0$; when $y_i(n)=1$, we define
\vspace{-0.5em}
\begin{equation}
	\vspace{-0.2em}
    \alpha_i(n) = \underbrace{\frac{1}{1 + e^{-\gamma_s\, \delta_i}}}_{\text{saturation factor}} \cdot \underbrace{\left[1 - \left(\frac{H_{i}(n)}{H_{i}^{\mathrm{thd}}(n)}\right)^{\!\lambda}\right]}_{\text{boundary-suppression factor}},
    \label{eq:confidence}
\end{equation}
where $\delta_i = H_{i}^{\mathrm{thd}}(n) - H_{i}(n)$, $\gamma_s$ controls saturation steepness, and $\lambda$ controls boundary suppression. The saturation factor outputs ${\sim}0.5$ at $\delta_i\approx 0$ and saturates toward $1$ as $\delta_i$ grows; the boundary-suppression factor tends to $0$ as the entropy/threshold ratio approaches $1$ and to $1$ otherwise. Their product yields a \emph{double-suppression} effect in the boundary region, driving the confidence to ${\sim}10^{-3}$, while genuine DDoS windows ($H_{i}(n)\approx 0.1$, $H_{i}^{\mathrm{thd}}(n)\approx 0.6$) yield values approaching $1$. We set $\gamma_s=20$, $\lambda=10$ in Table~\ref{tab:params}.

The reporting interface, per-source tracking, and consensus-suppression logic could be adapted to other detectors. For an ML/DL classifier, a calibrated attack probability or decision margin could serve as $\alpha_i(n)$. However, Equation~\eqref{eq:confidence}, the mean-threshold reference, and the threshold-recovery and floor rules rely on lower entropy indicating stronger attack evidence and must be redesigned.

\begin{algorithm}[!htbp]
	\caption{Edge-report consensus aggregation algorithm}
	\label{alg:consensus}
	\small \DontPrintSemicolon
	\KwIn{edge-report registry $\mathcal{R}_{\mathcal E}$, expiration time $\tau_{\mathrm{exp}}$}
	
	\KwOut{edge confidence $\alpha_{\mathcal E}(n)$, edge threshold $\bar{H}_{\mathcal E}^{\mathrm{thd}}(n)$}
	
	$\mathcal{R}_{\mathcal E}^{\mathrm{valid}}
	\gets
	\left\{
	r\in\mathcal{R}_{\mathcal E}
	\mid
	t_{\mathrm{now}}-t_r\le\tau_{\mathrm{exp}}
	\right\}$\;
	
	\lIf{$\mathcal{R}_{\mathcal E}^{\mathrm{valid}}=\emptyset$}
	{\Return $0.0,\;0.0$}
	
	$\alpha_{\mathcal E}(n)
	\gets
	\max_{r\in\mathcal{R}_{\mathcal E}^{\mathrm{valid}}}
	\alpha_r(n)$\;
	
	$\bar{H}_{\mathcal E}^{\mathrm{thd}}(n)
	\gets
	\frac{1}
	{|\mathcal{R}_{\mathcal E}^{\mathrm{valid}}|}
	\sum_{r\in\mathcal{R}_{\mathcal E}^{\mathrm{valid}}}
	H_r^{\mathrm{thd}}$\;
	
	\Return
	$\alpha_{\mathcal E}(n),\;
	\bar{H}_{\mathcal E}^{\mathrm{thd}}(n)$\;
\end{algorithm}

\subsection{Per-Source Tracking and Consensus Aggregation}\label{subsec:consensus}

The aggregation controller maintains an edge-report registry $\mathcal{R}_{\mathcal E}$ keyed by edge-controller IP, storing each edge controller's latest $(\alpha(n),H^{\mathrm{thd}}(n),\,t)$. Per-source storage prevents the information loss that a single-scalar aggregation (e.g., averaging) would incur. Before each fused decision, Algorithm~\ref{alg:consensus} computes the consensus pair $(\alpha_{\mathcal E}(n),\bar{H}_{\mathcal E}^{\mathrm{thd}}(n))$, where $\alpha_{\mathcal E}(n)$ is the \emph{maximum} confidence over valid entries (i.e., logical-OR semantics: any active edge alert keeps the global alert active); $\bar{H}_{\mathcal E}^{\mathrm{thd}}(n)$ is the \emph{mean} of valid thresholds, acting as a cross-domain normal baseline. If all entries expire the algorithm returns $(0,0)$, disabling the cross-domain branches. Note that entries older than $\tau_{\mathrm{exp}}=10$\,s are first dropped.

If an edge controller stops reporting, its latest entry remains eligible until $\tau_{\mathrm{exp}}$, allowing fusion to continue across transient reporting gaps; after expiration, it is discarded and the summary uses any remaining valid sources.

\subsection{Consensus-Referenced Cross-Domain Fusion}\label{subsec:fusion}

Given the local observation $(H_{j}(n),H_{j}^{\mathrm{thd}}(n),\alpha_j(n))$ and the consensus pair $(\alpha_{\mathcal E}(n),\bar{H}_{\mathcal E}^{\mathrm{thd}}(n))$ in each detection window, the aggregation controller produces the final decision through the following three complementary rules, which respectively address the root causes identified in Section~\ref{subsec:causes}.

\subsubsection{Rule 1: Cross-domain consensus suppression (addressing Root Cause I)} When the aggregation controller's preliminary local decision is $y_j(n)=1$ and all of the following hold simultaneously,
\begin{enumerate}[label=(C\arabic*)]
\item $\bar{H}_{\mathcal E}^{\mathrm{thd}}(n)>0$ (a valid edge report exists);
\item $\alpha_{\mathcal E}(n)=0$ (no edge controller is alarming);
\item $\alpha_j(n)<\gamma$ (the local confidence is below the suppression threshold $\gamma$),
\end{enumerate}
the decision is flipped to $\hat{y}(n)=0$. This rule precisely targets Root Cause~I: the attack has ended and edge controllers have recovered, but the aggregation controller produces a low-confidence local alarm due to OpenFlow statistics lag. We set $\gamma=0.15$, exploiting the ${\sim}10^{-3}$ depression that Equation~\eqref{eq:confidence} produces in the boundary region.

\subsubsection{Rule 2: Proactive threshold recovery (a weak version rule addressing Root Cause II)} When the edge consensus satisfies $\bar{H}_{\mathcal E}^{\mathrm{thd}}(n)>0$ and $\alpha_{\mathcal E}(n)=0$, the aggregation controller's EWMA is allowed to update with the current entropy value, even when the local decision remains $y_j(n)=1$. This ``controlled unlocking'' breaks the ``threshold locked at a low level'' deadlock, allowing the threshold to recover gradually toward the normal level.

\subsubsection{Rule 3: Edge-aware threshold-floor protection (a strong version rule addressing Root Cause II)} To prevent the EWMA threshold from collapsing at its source, an edge-aware dynamic floor is imposed after the threshold computation but before the traffic-adaptive freezing constraint:
\begin{equation}
    H_{j}^{\mathrm{thd}}(n) \gets \max\!\bigl(H_{j}^{\mathrm{thd}}(n),\;\min(\bar{H}_{\mathcal E}^{\mathrm{thd}},\;\phi)\bigr),
    \label{eq:floor}
\end{equation}
where $\phi$ is a preset upper limit that prevents the floor from being so high as to harm sensitivity. Since $\bar{H}_{\mathcal E}^{\mathrm{thd}}$ reflects the entropy level under normal network conditions (typically $0.6\sim0.7$), it serves as a dynamically updated healthy baseline. We set $\phi=0.5$, so that typical DDoS entropy values ($0.05\sim0.30$) remain well below the floor and reliably trigger detection.

\subsection{Joint Detection: The Complete Algorithm}\label{subsec:algorithm}

Algorithm~\ref{alg:joint} integrates all the components above. A key design choice is that {all cross-domain branches are gated by the unified \textit{cross-domain branch enabling condition} $\bar{H}_{\mathcal E}^{\mathrm{thd}}(n)>0$}. Since edge controllers do not receive reports from any other controller, their edge-report registry $\mathcal{R}_{\mathcal E}$ is always empty and $\alpha_{\mathcal E}(n)=\bar{H}_{\mathcal E}^{\mathrm{thd}}(n)=0$, so the cross-domain branches are naturally disabled. What remains on an edge controller is exactly the unmodified local detection logic, yielding a \emph{fully backward-compatible} incremental deployment. Table~\ref{tab:params} summarizes the configuration parameters used throughout.

 \begin{algorithm}[!htbp]
\caption{Aggregation-controller joint detection algorithm}
\label{alg:joint}
\small \DontPrintSemicolon
\KwIn{current-window data, an edge-report registry $\mathcal{R}_{\mathcal E}$, parameters $\gamma_s,\lambda,\gamma,\phi,\theta^{\mathrm{thd}},\tau_{\mathrm{exp}}$}
\KwOut{DDoS detection decision $\hat{y}(n) \in \{0,1\}$}
Obtain $(\alpha_{\mathcal E}(n), \bar{H}_{\mathcal E}^{\mathrm{thd}}(n))$   by Algorithm~\ref{alg:consensus}

Compute the current-window entropy $H_{j}(n)$ and base threshold $H_{j}^{\mathrm{thd}}(n)$ as in Equation~\eqref{eq:local_rule}\;
\If({ \tcp*[f]{threshold-floor protection}}){$\bar{H}_{\mathcal E}^{\mathrm{thd}}(n) > 0$}{
    $H_{j}^{\mathrm{thd}}(n) \gets \max(H_{j}^{\mathrm{thd}}(n),\;\min(\bar{H}_{\mathcal E}^{\mathrm{thd}}(n),\;\phi))$\;
}
\eIf{$H_{j}(n) < H_{j}^{\mathrm{thd}}(n)$}{
    Compute $\alpha_j$ via Equation~\eqref{eq:confidence}\;
    $\hat{y} \gets 1$\;
    \If(\tcp*[f]{cross-domain consensus suppression}){$\bar{H}_{\mathcal E}^{\mathrm{thd}}(n) > 0$ and $\alpha_{\mathcal E}=0$ and $\alpha_j<\gamma$}{
        $\hat{y} \gets 0$\;
    }
}{
    $\hat{y} \gets 0$\;
}
\If(\tcp*[f]{proactive threshold recovery}){$\bar{H}_{\mathcal E}^{\mathrm{thd}}(n) > 0$ and $\alpha_{\mathcal E}(n)=0$}{
    Update the EWMA using the current $H_{j}(n)$\;
}
\Return $\hat{y}$\;
\end{algorithm}
\vspace{-0.2em}

\begin{table}[!htbp]
\centering
\caption{Configuration parameters of the joint detector.}
\label{tab:params}
\renewcommand\arraystretch{1.1}
\vspace{-0.5em}
\begin{tabular}{clc}
\toprule
\textit{Parameter} & \textit{Description} & \textit{Value} \\
\midrule
$\theta^{\mathrm{thd}}$ & Dynamic feature-selection activation rate (pps) & 15{,}000 \\
$\gamma_s$ & Sigmoid steepness & 20 \\
$\lambda$ & Power-suppression exponent & 10 \\
$\gamma$ & Cross-domain suppression threshold & 0.15 \\
$\tau_{\mathrm{exp}}$ & Edge-report expiration time (s) & 10 \\
$\phi$ & Threshold-floor cap on $c_j$ & 0.5 \\
\bottomrule
\end{tabular}
\vspace{-2.0em}
\end{table}

\section{Experimental Evaluation}\label{sec:experiment}

\subsection{Setup and Controlled-Experiment Design}\label{subsec:exp_design}

The testbed runs on an Intel Core i9-12900K (16 cores, 128\,GB RAM) under Ubuntu 20.04, with Mininet v2.3.0~\cite{mininet} realizing the topology of Fig.~\ref{fig:topo_multi} and three independent Ryu v4.30~\cite{ryu2014ryu} instances acting as $c_1$, $c_2$, and $c_3$ over TCP sockets. Non-attacking hosts maintain dual-channel TCP/UDP background traffic; $h_1$ in $c_1$ domain and $h_8$ in $c_3$ domain launch high-rate SYN/ACK/UDP flooding against the victim $h_5$ in $c_2$ aggregation domain. Each detection window contains $N=400$ packets, and OpenFlow statistics are polled every 1\,s. Each run covers normal, attack, and post-attack transition phases; we perform 10 independent runs.

We design a paired controlled comparison on the same raw packet windows and ground-truth labels from all 10 runs. (i)~\emph{Baseline}---the legacy local DDoS detector~\eqref{eq:local_rule} is deterministically replayed without cross-domain information or active cross-domain rules (Table~\ref{tab:baseline}); (ii)~\emph{Joint detection}---$c_1$ and $c_3$ run the unmodified local detector and report $m_i(n)$ to $c_2$, which executes Algorithm~\ref{alg:joint}. The replayed joint predictions exactly match the online records. Ground-truth labels are aligned to detection windows (labeled positive when the attack fraction $\geq 0.25$); we report accuracy, precision, recall, F1 score, and FPR.

\subsection{Detection Performance on the Aggregation Controller}\label{subsec:c2_perf}

Table~\ref{tab:c2_results} reports the per-run performance of $c_2$ under joint detection, and Table~\ref{tab:improvement} quantifies the gains over the paired baseline. The FPR drops from 8.87\% to 1.96\% (a 77.9\% relative reduction), precision and F1 score improve by 14.40 and 7.85 percentage points, respectively, while recall decreases by only 0.69 percentage points---a favourable trade-off, since excessive false alarms cause \emph{alert fatigue} and erode analysts' trust.

\begin{table}[!htbp]
	\centering
	\caption{Aggregation controller $c_2$ under joint detection: 10 independent runs.}
	\label{tab:c2_results}
	\renewcommand\arraystretch{1.1}
	\vspace{-0.5em}
	\begin{tabular}{cccccc}
		\toprule
		\# & \textit{Accuracy} & \textit{Precision} & \textit{Recall} & \textit{F1} & \textit{FPR} \\
		\midrule
		1 & .9813 & .9470 & .9881 & .9671 & .0214 \\
		2 & .9720 & .8970 & 1.0000 & .9457 & .0371 \\
		3 & .9839 & .9471 & .9908 & .9685 & .0183 \\
		4 & .9831 & .9555 & .9833 & .9692 & .0170 \\
		5 & .9862 & .9701 & .9784 & .9742 & .0110 \\
		6 & .9893 & .9667 & .9902 & .9783 & .0109 \\
		7 & .9868 & .9624 & .9922 & .9771 & .0154 \\
		8 & .9764 & .9116 & .9949 & .9515 & .0291 \\
		9 & .9945 & .9847 & .9961 & .9904 & .0062 \\
		10 & .9787 & .9389 & .9966 & .9669 & .0294 \\
		\midrule
		\textbf{Mean} & \textbf{.9832} & \textbf{.9481} & \textbf{.9911} & \textbf{.9689} & \textbf{.0196} \\
		\bottomrule
	\end{tabular}
\end{table}
\begin{table}[!htbp]
	\centering
	\caption{Paired local baseline and joint detection on $c_2$ (mean of the same 10 runs).}
	\label{tab:improvement}
	\renewcommand\arraystretch{1.1}
	\vspace{-0.5em}
	\begin{tabular}{lccc}
		\toprule
		\textit{Metric} & \textit{Before / Baseline} & \textit{After / Joint} & \textit{Improvement} \\
		\midrule
		Accuracy  & 93.46\% & 98.32\% & $+4.87$\,pp \\
		Precision & 80.41\% & 94.81\% & $+14.40$\,pp \\
		Recall    & 99.79\% & 99.11\% & $-0.69$\,pp \\
		F1        & 89.04\% & 96.89\% & $+7.85$\,pp \\
		FPR       & 8.87\%  & 1.96\%  & $-77.9$\% \\
		\bottomrule
	\end{tabular}
	\vspace{-1.0em}
\end{table}

\begin{figure}[!htbp]
	\centering
	\vspace{-0.8em}
	\includegraphics[width=0.95\columnwidth]{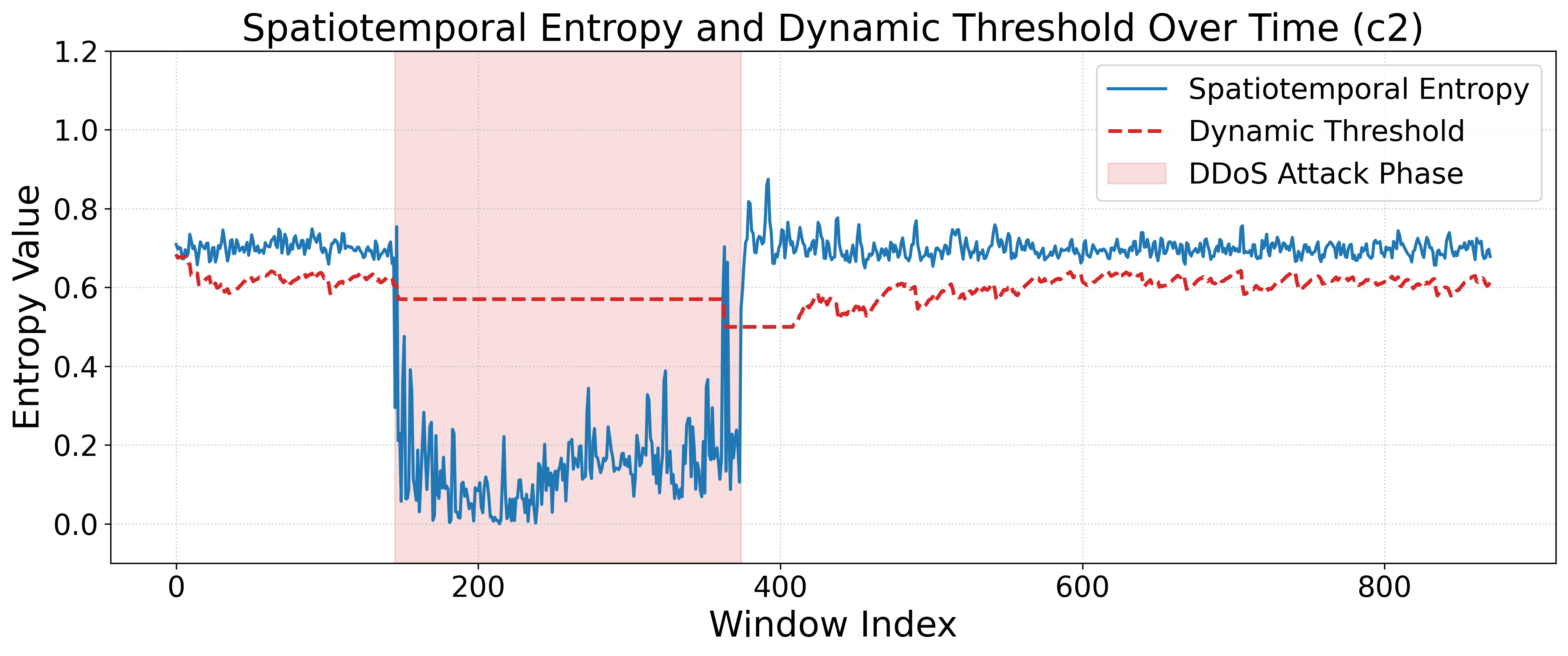}
	\caption{Temporal evolution of the local entropy and dynamic threshold on aggregation controller $c_2$. The edge-aware threshold floor prevents threshold collapse after the attack ends.}
	\vspace{-1.4em}
	\label{fig:entropy_threshold}
\end{figure}

Fig.~\ref{fig:entropy_threshold} further shows, for one typical run, the temporal evolution of the local entropy and dynamic threshold on $c_2$. In the post-attack transition phase, an unconstrained EWMA threshold keeps decaying as low entropy values are absorbed into the update; once the edge-aware floor is enabled, the threshold is anchored near the edge-consensus level, avoiding the threshold collapse identified in Root Cause~II---mechanistically consistent with the FPR reduction in Table~\ref{tab:improvement}.

\subsection{Zero-Impact Verification on Edge Controllers}\label{subsec:edge_isolation}

The proposed method does not alter the local entropy, dynamic-threshold, or decision computations at $c_1$ and $c_3$: the edge controllers only compute and report confidence and threshold values to $c_2$, while all cross-domain rules execute at $c_2$. Consistent with this design, the 10-run paired replay leaves their Accuracy, F1, and FPR unchanged ($c_1$: 98.84\%, 98.84\%, and 0.93\%; $c_3$: 98.40\%, 98.53\%, and 3.33\%, respectively), confirming that the improvement is confined to aggregation-side detection.

\subsection{Communication and Processing Overhead}\label{subsec:overhead}

We quantify overhead over the same 10 current joint-detection runs. A framed report comprises the compact JSON payload and its 4-byte length header. Edge-side instrumentation records serialization cost and application-layer throughput. At $c_2$, report handling includes accept-to-handler dispatch, JSON decoding, and per-source registry update, while socket receive waiting is excluded; consensus and fusion-rule time is measured once per detection window. Each event metric is first averaged within a run and then reported as the mean $\pm$ sample standard deviation of the 10 run-level means. The results are limited to application-layer communication and controller-side processing rather than wire-level TCP/IP bandwidth.

\begin{table}[!htbp]
\centering
\caption{Application-layer and controller-side overhead (mean $\pm$ sample standard deviation of 10 run-level means).}
\label{tab:overhead}
\renewcommand\arraystretch{1.1}
\setlength{\tabcolsep}{4pt}
\footnotesize
\begin{tabular}{lcc}
\toprule
\textit{Metric} & \textit{Mean $\pm$ SD} & \textit{Unit} \\
\midrule
\multicolumn{3}{l}{\textit{Communication}} \\
Framed report size & $53.13 \pm 0.32$ & B/report \\
Edge serialization & $16.26 \pm 0.41$ & $\mu$s/report \\
Reporting rate per edge & $3.40 \pm 0.28$ & msg/s \\
Aggregate application-layer data rate & $2.89 \pm 0.24$ & kb/s \\
\midrule
\multicolumn{3}{l}{\textit{Processing at $c_2$}} \\
Report handling & $2{,}332.73 \pm 330.39$ & $\mu$s/report \\
Consensus and fusion-rule processing & $14.38 \pm 2.39$ & $\mu$s/window \\
Amortized collaboration processing & $1{,}977.46 \pm 297.45$ & $\mu$s/window \\
\bottomrule
\end{tabular}
\end{table}

As shown in Table~\ref{tab:overhead}, the framed report size is $53.13\pm0.32$\,B and each edge sends $3.40\pm0.28$ reports/s, for an aggregate application-layer rate of $2.89\pm0.24$\,kb/s. Edge serialization requires $16.26\pm0.41\,\mu$s per report. At $c_2$, report handling averages $2.33$\,ms per report; its pooled median is $269.58\,\mu$s and its 95th percentile is $7.61$\,ms, showing that thread-dispatch delay creates a right-skewed tail. Consensus and fusion rules require only $14.38\pm2.39\,\mu$s per window. For each run, dividing the sum of all measured report-handling and fusion-processing time by the number of $c_2$ fusion windows gives an amortized cost of $1.98\pm0.30$\,ms per window; the campaign contains $0.84\pm0.03$ successfully received reports per fusion window.

\subsection{Parameter Sensitivity}\label{subsec:sensitivity}

We evaluate parameter sensitivity by deterministic one-at-a-time replay of the 10 current $c_2$ traces. The replay preserves the observed window and report order, and the default configuration reproduces the online predictions exactly. We vary sigmoid steepness $\gamma_s\in\{10,20,30\}$, power exponent $\lambda\in\{5,10,15\}$, suppression threshold $\gamma\in\{0.10,0.15,0.20\}$, and threshold-floor cap $\phi\in\{0.4,0.5,0.6\}$ while holding all other parameters fixed. Table~\ref{tab:sensitivity} reports the mean and sample standard deviation across runs.

\begin{table}[!htbp]
\centering
\caption{One-at-a-time sensitivity on $c_2$ (mean $\pm$ sample SD over 10 runs; $^\ast$ denotes the default).}
\label{tab:sensitivity}
\renewcommand\arraystretch{1.05}
\setlength{\tabcolsep}{2.5pt}
\scriptsize
\begin{tabular}{ccccc}
\toprule
\textit{Parameter} & \textit{Value(s)} & \textit{Recall} & \textit{F1} & \textit{FPR} \\
\midrule
$\gamma_s$ & 10, $20^\ast$, 30 & $.9911\pm.0065$ & $.9689\pm.0129$ & $.0196\pm.0097$ \\
$\lambda$ & 5, $10^\ast$, 15 & $.9911\pm.0065$ & $.9689\pm.0129$ & $.0196\pm.0097$ \\
$\gamma$ & 0.10, $0.15^\ast$, 0.20 & $.9911\pm.0065$ & $.9689\pm.0129$ & $.0196\pm.0097$ \\
$\phi$ & 0.4 & $.9872\pm.0174$ & $.9671\pm.0133$ & $.0194\pm.0099$ \\
$\phi$ & $0.5^\ast$ & $.9911\pm.0065$ & $.9689\pm.0129$ & $.0196\pm.0097$ \\
$\phi$ & 0.6 & $.9954\pm.0043$ & $.9699\pm.0135$ & $.0205\pm.0097$ \\
\bottomrule
\end{tabular}
\vspace{-1.0em}
\end{table}

The confidence-shaping parameters $\gamma_s$ and $\lambda$ alter continuous confidence values but do not change a prediction in this cohort; varying $\gamma$ is likewise decision-invariant. Diagnostic replay shows that the consensus-suppression and proactive-recovery conditions are not activated in these traces, so this invariance should be interpreted as conditional branch inactivity rather than global insensitivity. In contrast, $\phi$ directly changes the threshold trajectory: values 0.4 and 0.6 alter 10 and 16 of the 8,833 evaluated decisions, respectively. Increasing $\phi$ from 0.5 to 0.6 raises recall from 99.11\% to 99.54\%, with FPR increasing from 1.96\% to 2.05\%; the default $\phi=0.5$ provides a balanced operating point.

\subsection{Residual FPs, Positioning, and Limitations}\label{subsec:fp_compare}

The current joint cohort retains 12.7 false-positive windows per run on average. Two mechanisms can contribute to these residual errors: boundary windows in which attack and normal traffic coexist below the 0.25 labeling threshold, and post-attack high-rate windows in which dynamic feature selection remains active after traffic has begun to recover. Raising $\theta^{\mathrm{thd}}$ can reduce the latter at the cost of delayed early-attack adaptation.

Existing multi-controller schemes~\cite{giotis2014leveraging,Entropy_dstIP_2015,yu2021cooperative,tayfour2021collaborative,wang2023ccguard} focus on information sharing across controllers; our problem is the aggregation-side bias itself. Table~\ref{tab:comparison} positions our work and clarifies the research gap rather than implying superiority.

\begin{table}[!htbp]
\centering
\caption{Methodological positioning of multi-controller cooperative DDoS detection.}
\label{tab:comparison}
\renewcommand\arraystretch{1.1}
\setlength{\tabcolsep}{4pt}
\footnotesize
\vspace{-0.5em}
\begin{tabular}{lccc}
\toprule
\textit{Scheme} & \textit{Information / app. direction} & \textit{Bias$^\sharp$} & \textit{Edge$^\S$} \\
\midrule
Giotis~\cite{giotis2014leveraging}        & Flow stats / bi-dir.      & No  & Mod. \\
Wang~\cite{Entropy_dstIP_2015}            & Entropy / bi-dir.         & No  & Mod. \\
Yu~\cite{yu2021cooperative}               & Coarse alarm / bi-dir.    & No  & Mod. \\
Tayfour~\cite{tayfour2021collaborative}   & Rule msg / broadcast      & No  & Mod. \\
\textbf{This paper}                       & Conf.\ + thd.\ / edge$\to$agg. & \textbf{Yes} & \textbf{Untouched} \\
\bottomrule
\end{tabular}
\\\footnotesize \textit{Bias$^\sharp$}: explicitly models aggregation-side bias. \textit{Edge$^\S$}: edge-controller is modified (Mod.) or untouched.
\vspace{-1.0em}
\end{table}

This work has several limitations. The current evaluation focuses on a three-controller chain topology, high-rate SYN/ACK/UDP flooding, and inter-controller links satisfying assumption~\textbf{(A2)}. The current traces do not activate the suppression and proactive-recovery branches, so the observed decision invariance for $\gamma_s$, $\lambda$, and $\gamma$ is specific to this traffic cohort. Larger or non-chain topologies, low-rate attacks, and partial or delayed report delivery are left for future evaluation.

The current design assumes trusted controllers; extending the maximum-confidence and mean-threshold rules to malicious reports would require Byzantine-robust aggregation and authenticated reporting with integrity and freshness checks. These extensions are left for future work.

\section{Conclusion}\label{sec:conclusion}

This paper characterized the systematic false-positive problem of an entropy-based detector deployed on the aggregation controller of a multi-controller SDN as an \emph{aggregation bias} caused by OpenFlow statistics lag and unconstrained EWMA threshold drift, and proposed a cross-domain confidence-fusion joint-detection method comprising a nonlinear confidence metric, per-source edge consensus, consensus-referenced fusion, and an edge-aware threshold floor. On a three-controller, 24-host Mininet testbed, the method reduces the aggregation FPR from 8.87\% to 1.96\% and raises F1 from 89.04\% to 96.89\%, with edge-controller performance unchanged. Controller-side measurements further show an application-layer reporting rate of 2.89\,kb/s and an amortized aggregation overhead of 1.98\,ms per window.


\bibliographystyle{IEEEtran}
\bibliography{references}
\end{document}